\documentclass[sigconf]{acmart}
\usepackage{xcolor, graphicx}
\usepackage[export]{adjustbox}
\usepackage{subcaption}

\AtBeginDocument{%
	\providecommand\BibTeX{{%
			\normalfont B\kern-0.5em{\scshape i\kern-0.25em b}\kern-0.8em\TeX}}}

\setcopyright{acmcopyright}
\copyrightyear{2020}
\acmYear{2020}
\acmDOI{10.1145/1122445.1122456}

\acmConference[CODS-COMAD '21]{CODS-COMAD '21: ACM India Joint International Conference on Data Science \& Management of Data}{January 02--04, 2021}{IIIT Bangalore, India}
\acmBooktitle{CODS-COMAD '21: ACM India Joint International Conference on Data Science \& Management of Data, January 02--04, 2021, IIIT Bangalore, India}
\acmPrice{15.00}
\acmISBN{978-1-4503-XXXX-X/18/06}

\begin{document}
\fancyhead{}

\title{How Have We Reacted To The COVID-19 Pandemic? Analyzing Changing Indian Emotions Through The Lens of Twitter}


\author{Rajdeep Mukherjee}
\affiliation{\institution{IIT Kharagpur, India}}
\email{rajdeep1989@iitkgp.ac.in}

\author{Sriyash Poddar}
\authornote{Both authors contributed equally to this research.}
\email{poddarsriyash@iitkgp.ac.in}
\affiliation{\institution{IIT Kharagpur, India}}

\author{Atharva Naik}
\authornotemark[1]
\email{atharvanaik2018@iitkgp.ac.in}
\affiliation{\institution{IIT Kharagpur, India}}

\author{Soham Dasgupta}
\affiliation{\institution{Mallya Aditi International School, Bangalore, India}}
\email{sohamdasgupta91@gmail.com}

\renewcommand{\shortauthors}{Rajdeep Mukherjee, et al.}

\begin{abstract}
	Since its outbreak, the ongoing COVID-19 pandemic has caused unprecedented losses to human lives and economies around the world. As of 18th July 2020, the World Health Organization (WHO) has reported more than 13 million confirmed cases including close to 600,000 deaths across 216 countries and territories. Despite several government measures, India has gradually moved up the ranks to become the third worst-hit nation by the pandemic after the US and Brazil, thus causing widespread anxiety and fear among her citizens. As majority of the world's population continues to remain confined to their homes, more and more people have started relying on social media platforms such as Twitter for expressing their feelings and attitudes towards various aspects of the pandemic. With rising concerns of mental well-being, it becomes imperative to analyze the dynamics of public affect in order to anticipate any potential threats and take precautionary measures. Since affective states of human mind are more nuanced than meager binary sentiments, here we propose a deep learning-based system to identify people's emotions from their tweets. We achieve competitive results on two benchmark datasets for multi-label emotion classification. We then use our system to analyze the evolution of emotional responses among Indians as the pandemic continues to spread its wings. We also study the development of salient factors contributing towards the changes in attitudes over time. Finally, we discuss directions to further improve our work and hope that our analysis can aid in better public health monitoring.
\end{abstract}

\keywords{Covid-19, Pandemic, Fine-grained Sentiment Analysis, Multi-label Emotion Classification, Tweets}

\maketitle

\section{introduction}\label{intro}
The coronavirus disease 2019, or the COVID-19, caused by the SARS-Cov-2 virus, has subsequently taken the shape of a pandemic since it's outbreak in late December 2019, causing enormous losses to societies and economies around the world \cite{Kleinberg2020MeasuringEI}. Although the enforcement of several protective measures such as \textit{lockdown, social distancing,} etc. have had positive effects on containing the spread of the virus, the loneliness caused by self-confinement and reduced access to family, friends, and other social support systems have caused severe psychological threats to the physical and mental well-being of people around the globe \cite{Ammar2020EffectsOH, Hiremath2020COVID1I, Gao2020MentalHP}. While dealing with such unprecedented challenges, as more and more people express their emotions and opinions towards various aspects of the pandemic through social media platforms such as Twitter, Facebook, etc., it becomes essential to monitor the dynamics of public affect from such user-generated textual content in order to understand general concerns and anticipate any potential threats.

Human-written texts such as tweets reflect an author's affective state or thought process. Although emotions and sentiments are closely related, merely classifying the subjective content of a tweet into positive, negative or neutral sentiment categories may not be sufficient to understand a person's true feelings. Under trying situations such as the COVID-19 pandemic, emotions can be even more complicated \cite{Yang2020SenWaveMT}. A person might feel optimistic about vaccine development or thankful towards the doctors or even both while still being positive. Therefore, in order to capture people's emotions from their tweets, we build a \textit{transformer-based} \cite{Vaswani2017AttentionIA} supervised multi-label emotion classifier that achieves state-of-the-art results on two benchmark datasets, \textit{AIT} \cite{Mohammad2018SemEval2018T1}, and \textit{SenWave} \cite{Yang2020SenWaveMT}.

With record no. of cases being reported daily and more than 1 million confirmed cases to date, India has now become the third worst-hit nation by the pandemic after the US and Brazil. We, therefore use our classifier, trained on COVID-19-specific \textit{SenWave} dataset, to monitor the evolution of emotional attitudes among Indians towards the ongoing pandemic by analyzing tweets posted between March 1st 2020 and July 5th 2020. We also study the development of salient factors or triggers contributing towards the changes in emotions over time. We hope that our proposed system can aid the concerned authorities in real-time identification of peoples' mental health conditions and concerns from their social media posts and make timely interventions in fighting the global crisis.

\section{Related Work}\label{related}
Studies based on social media data around the COVID-19 pandemic have seen a steep surge over the past few months. Prior works \cite{Li2020AnalyzingCO, Ahmed2018MoralPT, Leis2019DetectingSO} have stressed the need for automatic detection and monitoring of public affects from user-generated content, especially during trying times such as the current pandemic. Although recent benchmark annotated datasets \cite{Mohammad2017WASSA2017ST, Mohammad2018SemEval2018T1, Demszky2020GoEmotionsAD} have facilitated the training and development of supervised deep learning-based classifiers for automated sentiment analysis and emotion detection, their results are not directly applicable to COVID-19 analysis \cite{Yang2020SenWaveMT} due to considerable differences in vocabulary between the training and testing domains. Majority of recent works around COVID-19 such as \cite{Kabir2020CoronaVisAR, Xue2020MachineLO} have thus relied on feature-engineering based methods or supervised methods with limited training data \cite{Kleinberg2020MeasuringEI}. As the pandemic continues to spread its wings, we draw our motivation from \cite{Mohammad2018SemEval2018T1, Yang2020SenWaveMT, Li2020AnalyzingCO} for building a multi-label emotion classifier and studying the evolution of public emotions towards the COVID-19 from an Indian perspective, through the lens of Twitter.
\section{Datasets}\label{datasets}
\begin{itemize}
	\item \textbf{AIT :} The \textit{Affect in Tweets Dataset} (or \textit{AIT} for short) was created in \cite{Mohammad2018SemEval2018T1}, as part of SemEval-2018 Task 1. This non-COVID-19 dataset consists of 7724 English tweets, each labeled with either \textit{neutral or no emotion} or one or more of the 11 emotions, \textit{anger, anticipation, disgust, fear, joy, love, optimism, pessimism, sadness, surprise}, and \textit{trust}.
	\item \textbf{SenWave :} We consider the 10K English tweets from the largest fine-grained annotated Covid-19 tweets dataset created by the authors in \cite{Yang2020SenWaveMT}. Here, each tweet is labeled with one or more of the 10 emotions, \textit{optimistic, thankful, empathetic, pessimistic, anxious, sad, annoyed, denial} (towards conspiracy theories), \textit{official report}, and \textit{joking}.
	\item \textbf{Twitter\_IN :} We use \textit{version 17} of the COVID-19 Twitter chatter dataset released by the authors in \cite{Banda2020ALC}. \textit{Twitter\_IN} consists of around 1.4 lac English tweets from India posted between January 25th 2020 and July 4th 2020.
\end{itemize}
The version of \textit{SenWave}, obtained from the authors, contains an additional label \textit{surprise}. Although we use all the 11 categories to train our models, however, we leave out \textit{surprise} (since not documented in \cite{Mohammad2018SemEval2018T1}) and official (since not an affective state) from further consideration while performing our analysis with \textit{Twitter\_IN}.

\section{Model Description}\label{model}
\subsection{Data Preprocessing}\label{data_preprocess}
We preprocessed the raw tweets in all the datasets, first by removing the \textit{mentions} (@user), and URLs. We then filtered out noisy symbols such as "RT" (retweet). Since \textit{hashtags} can provide meaningful semantics, we removed the \textit{\#} from the \textit{\#word} and kept the \textit{word}. We used a \textit{mapping} \footnote{https://en.wikipedia.org/wiki/List\_of\_emoticons/} between emoticons with their meanings and the \textit{emoji}\footnote{https://pypi.org/project/emoji/} package to respectively replace emoticons and emojis appearing in the tweets with their semantic phrases. We then used a \textit{slang\_translator} \footnote{https://github.com/rishabhverma17/sms\_slang\_translator} to convert online \textit{slangs}, commonly used in short text message conversations, into their actual phrases. For eg: "CUL8R" gets replaced by "see you later". We further expand the contractions using \textit{contractions}\footnote{https://pypi.org/project/contractions/}. Finally, we filtered out punctuations, line breaks, tabs and redundant blank characters, etc. since they do provide any relevant lexical or semantic information.

\subsection{Architecture}\label{architecture}
\begin{figure}[]
	\centering
	\includegraphics{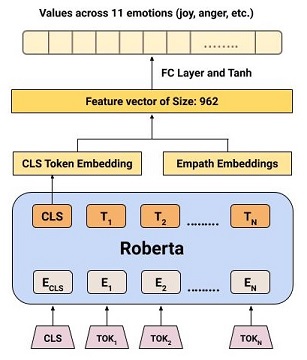}
	\caption{Model Architecture}
	\label{fig:model}
\end{figure}
We used RoBERTa-Base \cite{Liu2019RoBERTaAR} as the backbone of our multi-label emotion classifier. A tweet $ t $ to be classified is first broken down by the \textit{RobertaTokenizer} into its constituent tokens. A special token [CLS] is appended to obtain the final sequence $ [CLS], {t_{1}, t_{2}, ... , t_{n}} $, which is then passed through 12 layers of attention-based \textit{transformers} to obtain a 768-dimensional contextualized representation of the tweet, $ h_{CLS} $. Motivated by prior works \cite{Mohammad2018SemEval2018T1}, we simultaneously obtain a 194-dimensional feature vector derived from affect lexicons using \textit{Empath} \cite{Fast2016EmpathUT}. We append the two to obtain a 962-dimensional vector which is then passed through a fully connected layer with 11 output neurons. At each neuron, we use the \textit{tanh} activation function with a threshold of 0.33 to finally predict the presence/absence of one of the 11 emotions. hereafter, we refer to this model as EC\textsubscript{ROBERTA}. A variant of our model with BERT-Base \cite{Devlin2019BERTPO} as the backbone is referred to as EC\textsubscript{BERT}. Among our non-BERT variants, EC\textsubscript{CNN} consists of five parallel 1-D convolutions, each with different kernel sizes ranging from 2-6 and \textit{ReLU} activation function, followed by 1-D max-pooling layers. All five outputs are merged to obtain a single representation vector for the tweet. EC\textsubscript{LSTM} consists of a single layer of unidirectional LSTM cells with 256 hidden units. Hidden state from the last LSTM cell is considered as the tweet representation vector. For both the above models, the obtained tweet vector is passed through a hidden layer consisting of 128 neurons followed by a fully connected layer with 11 output neurons. Labels at each neuron are predicted with \textit{sigmoid} activation function.
\section{Results and Analysis}\label{results}
\begin{table}
	\small
	\caption{Comparison of Results on \textit{AIT}.}
	\label{tab:ec_comparison_ait}
	\centering
	\begin{tabular}{lccc}
		\hline
		\textbf{Methods} & \textbf{Jaccard Acc.} & \textbf{F1-Macro} & \textbf{F1-Micro} \\ \hline
		NTUA-SLP \cite{Baziotis2018NTUASLPAS} & 0.588 & 0.528 & 0.701 \\
		Current Leader & 0.594 & 0.565 & 0.704 \\
		\hline
		EC\textsubscript{CNN} (Word2Vec) & 0.475 & 0.409 & 0.601  \\
		EC\textsubscript{CNN} (Glove) & 0.479 & 0.438 & 0.609  \\
		EC\textsubscript{CNN} (Glove + Empath) & 0.489 & 0.445 & 0.613  \\
		EC\textsubscript{LSTM} (Glove + Empath) & 0.509 & 0.463 & 0.625 \\
		EC\textsubscript{BERT} & 0.573 & 0.562 & 0.690  \\
		EC\textsubscript{ROBERTA} & \textbf{0.594} & \textbf{0.578} & \textbf{0.704} \\
		\hline		
	\end{tabular}
\end{table}

\begin{table}
	\small
	\caption{Comparison of Results on \textit{SenWave}.}
	\label{tab:covid_comparison}
	\centering
	\begin{tabular}{lcccccc}
		\hline
		\textbf{Methods} & \textbf{Acc.} & \textbf{J.Acc.} & \textbf{F1-Ma.} & \textbf{F1-Mi.} & \textbf{LRAP} & \textbf{H.Loss} \\ 
		\hline
		SenWave \cite{Yang2020SenWaveMT} & \textbf{0.847} & 0.495 & 0.517 & 0.573 & 0.745 & \textbf{0.153} \\
		EC\textsubscript{BERT} & 0.830 & 0.479 & 0.530 & 0.590 & 0.747 & 0.170 \\
		EC\textsubscript{ROBERTA} & 0.838 & \textbf{0.499} & \textbf{0.554} & \textbf{0.609} & \textbf{0.781} & 0.162 \\
		\hline		
	\end{tabular}
\end{table}

\subsection{Evaluation Metrics}\label{metrics}
Apart from \textit{Macro-F1} and \textit{Micro-F1}, both \cite{Mohammad2018SemEval2018T1} and \cite{Yang2020SenWaveMT} consider \textit{Jaccard Accuracy} as the principle metric to evaluate the multi-label accuracy of the models designed for multi-label emotion detection. \cite{Yang2020SenWaveMT} additionally considers \textit{Label Ranking average precision} (or LRAP), \textit{Hamming Loss}, and a relaxed accuracy mesaure called \textit{Weak Accuracy} or just "Accurcay" to evaluate the models.

\subsection{Multi-label Emotion Classification}\label{ec}
Table 1 compares our results with those of the baselines for the task of multi-label emotion classification when the models are trained on the \textit{AIT} dataset. An ablation study performed on EC\textsubscript{CNN} demonstrates the advantage of using richer word representations such as \textit{Glove} \cite{Pennington2014GloveGV} over \textit{Word2Vec} \cite{Mikolov2013DistributedRO}. It further establishes the importance of using affect representations from text such as \textit{Empath} embeddings for sentiment analysis tasks. Among our non-BERT variants, EC\textsubscript{LSTM} performs the best. We obtain our best results with EC\textsubscript{ROBERTA}, when trained end-to-end for 5 epochs with \textit{AdamW} optimizer \cite{Loshchilov2019DecoupledWD} and a learning rate of $ 2e-5 $. As observed, we comfortably outperform the scores of \textit{NTUA-SLP} \cite{Baziotis2018NTUASLPAS}, which achieved the top rank for this subtask as part of the SemEval 2018 Task 1. With a better F1-Macro score, we narrowly surpass the results of the current top-ranked system on the leaderboard for this subtask, with the competition currently running in its post-evaluation phase. As can be observed from Table 2, EC\textsubscript{ROBERTA}, when trained on \textit{SenWave} dataset further achieves state-of-the-art results on four out of six metrics, including \textit{Jaccard Accuracy}, when compared with \cite{Yang2020SenWaveMT}.

\subsection{The Evolution of Public Emotions and Contributing Factors}\label{analysis}
\begin{figure}
	\centering
	\includegraphics[width=\columnwidth]{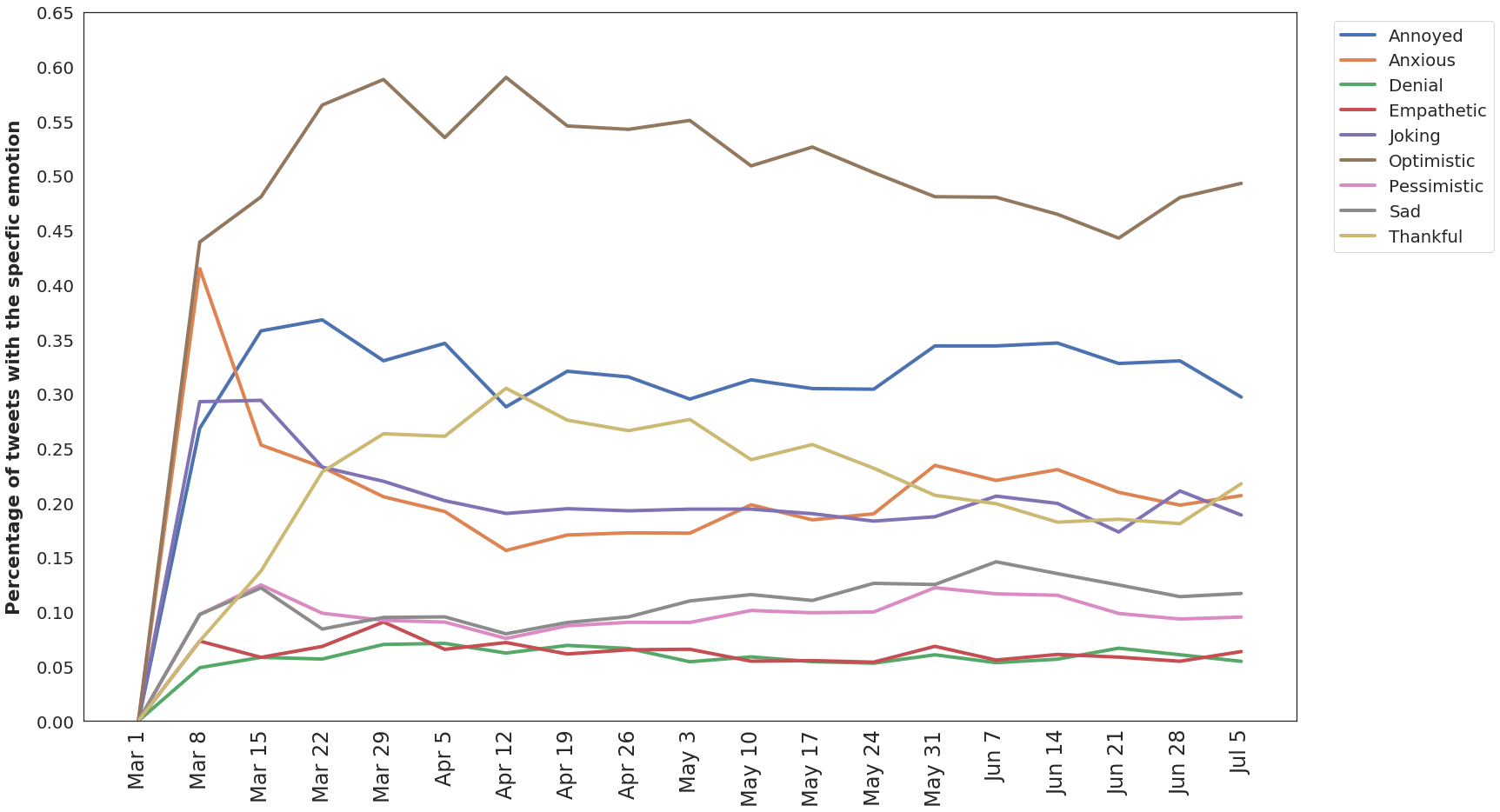}
	\caption{Weekly percentage distribution of various public emotions on Twitter.}
	\label{fig:percent-wise}
\end{figure}

\begin{figure}
	\centering
	\includegraphics[width=\columnwidth]{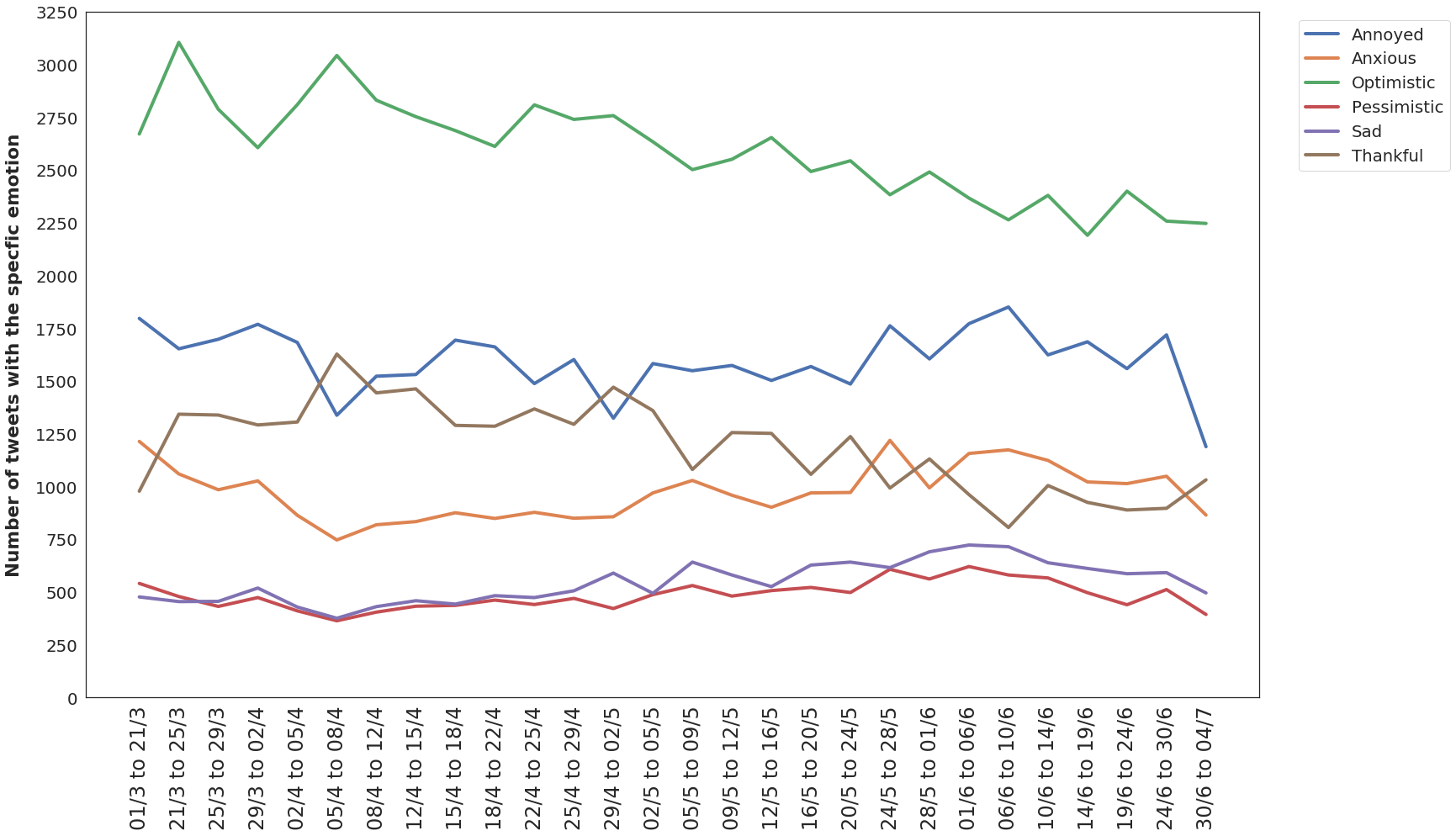}
	\caption{Count distribution of tweets containing a specific emotion per 5000 tweets posted.}
	\label{fig:chunk-wise}
\end{figure}

Owing to the fact that there were only three reported cases of COVID-19 in India before March 2020, and there exist very few tweets in \textit{Twitter\_IN} for the months of January and February, we consider all tweets posted on or after Match 1st 2020 for our analysis. We first use our EC\textsubscript{ROBERTA} classifier, trained on \textit{SenWave} dataset, to predict the emotions for all tweets under consideration from the \textit{Twitter\_IN} dataset. Few tweets with their predicted emotions are listed in Table 3. Figure \ref{fig:percent-wise} shows the percentage distribution of tweets, on a weekly basis, containing one or more from a total of nine emotions. We observe that \textit{annoyed} and \textit{thankful} show contrasting trends. First and second peaks of annoyance correspond with the declaration of nationwide lockdown and the \textit{Tablighi Jamaat} gatherings respectively. Although, people initially showed their gratitude towards the doctors and health workers for their efforts in dealing with such unprecedented challenges, the sense of thankfulness gradually declined as people started to feel more helpless and anxious with growing no. of cases each day. Here, we highlight that the no. of tweets posted greatly varies across weeks. In Figure \ref{fig:chunk-wise}, we therefore create unequal-sized bins, each however containing exactly 5000 tweets (except the last one, which contains around 4,300 tweets), and observe the trends for six most relevant emotions. We observe that the trends of \textit{optimism} and \textit{pessimism} complement each other well as people gradually start to adjust themselves with the new normal of life.

\begin{table*}
	\caption{Few Examples of Single and Multi-label Predictions}
	\label{tab:prediction_examples}
	\centering
	\begin{tabular}{p{0.75\textwidth}c}
		\hline
		\makebox[0.75\textwidth]{\textbf{Tweet}} & \textbf{Predicted Labels} \\
		\hline
		\multicolumn{2}{l}{\vspace{0.05 in}\textbf{Single Label}} \\ 
		\hline
		This is the time to fight Covid19 at present but some intelligent Generals are focusing on war and terrorism & Annoyed \\
		\hline
		let us stay together and fight against Corona VirusCoronavirusPandemic Lockdown2 & Optimistic \\
		\hline
		it is a serious matter and thousands of students live are in danger due to increasing cases of COVID19 Cancel exams and promote the students on moderation policy & Anxious \\
		\hline
		\multicolumn{2}{l}{\vspace{0.05 in}\textbf{Multiple Labels}} \\ 
		\hline
		Media is so obsessed with a particular community that they even misspell coronavirus & Annoyed, Joking, Surprise \\
		\hline
		Very nice Measures taken for The Development Of Economy keeping in view the safest side & Thankful, Optimistic \\
		\hline
		The first Covid 19 positive from Meghalaya Dr John Sailo Rintathiang passed away early this morning Sailo 69 who was also the owner of Bethany hospital was tested positive on April 13 2020 & Sad, Official Report \\
		\hline				
	\end{tabular}
\end{table*}

\begin{table*}[]
	\caption{Top contributing factors or aspects affecting different emotions on Twitter regarding COVID-19 pandemic.}
	\label{tab:top10_terms}
	\centering
	\begin{tabular}{lc}
		\hline
		\textbf{Emotion} & \textbf{Top 10 aspects} \\ 
		\hline
		\textbf{Anxious} & family, symptom, test, treatment, rate, risk, mask, spread, zone, assault \\
		\textbf{Annoyed} & govt, politics, death, news, religion, jamaat, work, China, assault, border \\
		\textbf{Sad} & lockdown, death, distancing, life, family, economy, village, doctor, worker, school \\
		\textbf{Pessimistic} & price, business, infection, demise, peak, curve, communalism, war, situation, transport \\
		\textbf{Optimistic} & initiative, opportunity, measure, arogyasetuapp, IndiaFightsCorona, stayhome, contribution, change, support, action \\
		\textbf{Thankful} & doctor, service, staff, nurse, app, fund, assistance, leadership, IndiaFightsCorona \\
		\hline		
	\end{tabular}
\end{table*}

\begin{figure*}
	\centering
	\begin{subfigure}[b]{.45\linewidth}
		\textcolor{lightgray}{\includegraphics[width=\linewidth, frame]{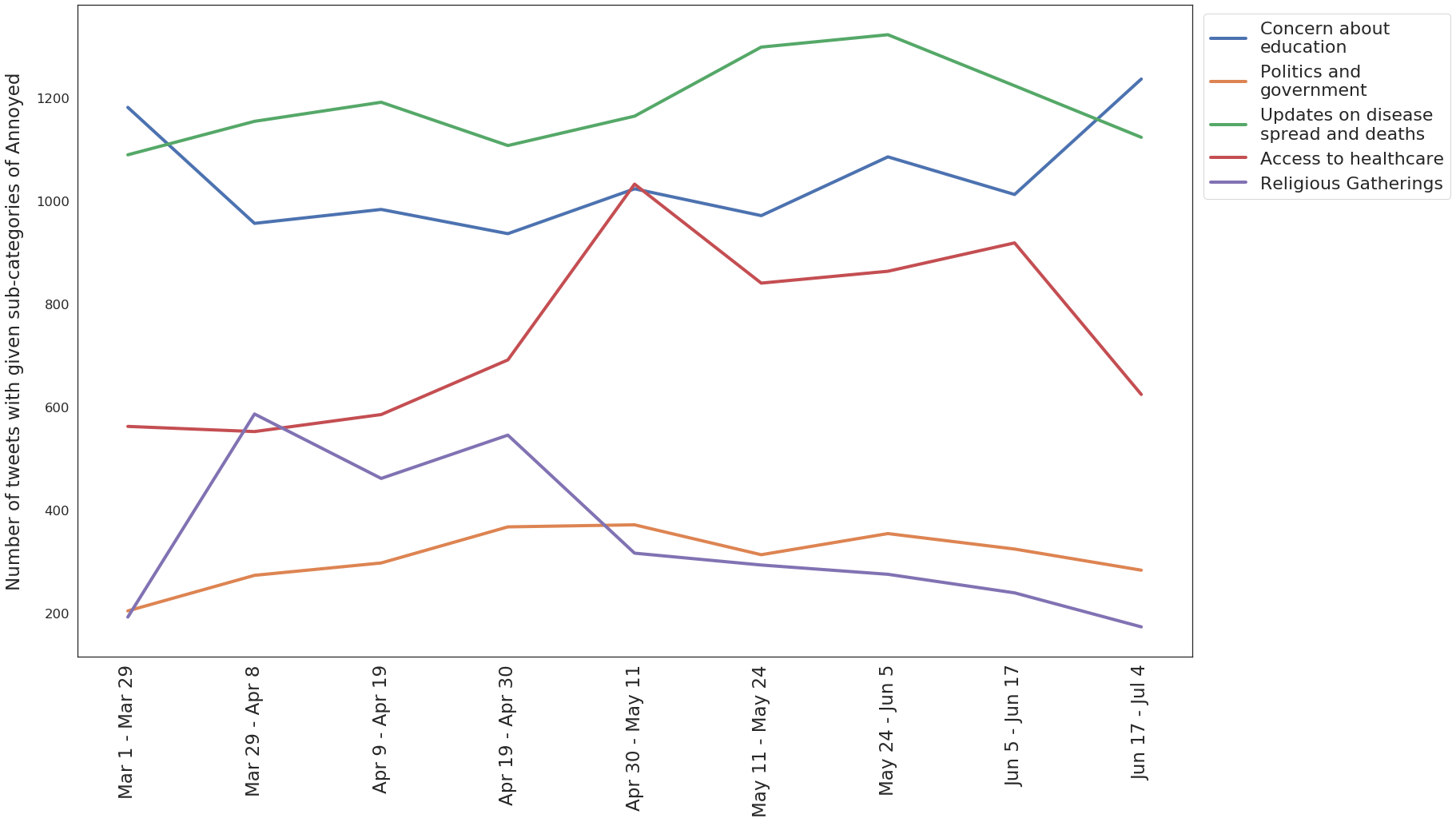}}
		\caption{Annoyed}
		\label{fig:annoyed_trend}
	\end{subfigure}
	\begin{subfigure}[b]{.45\linewidth}
		\textcolor{lightgray}{\includegraphics[width=\linewidth, frame]{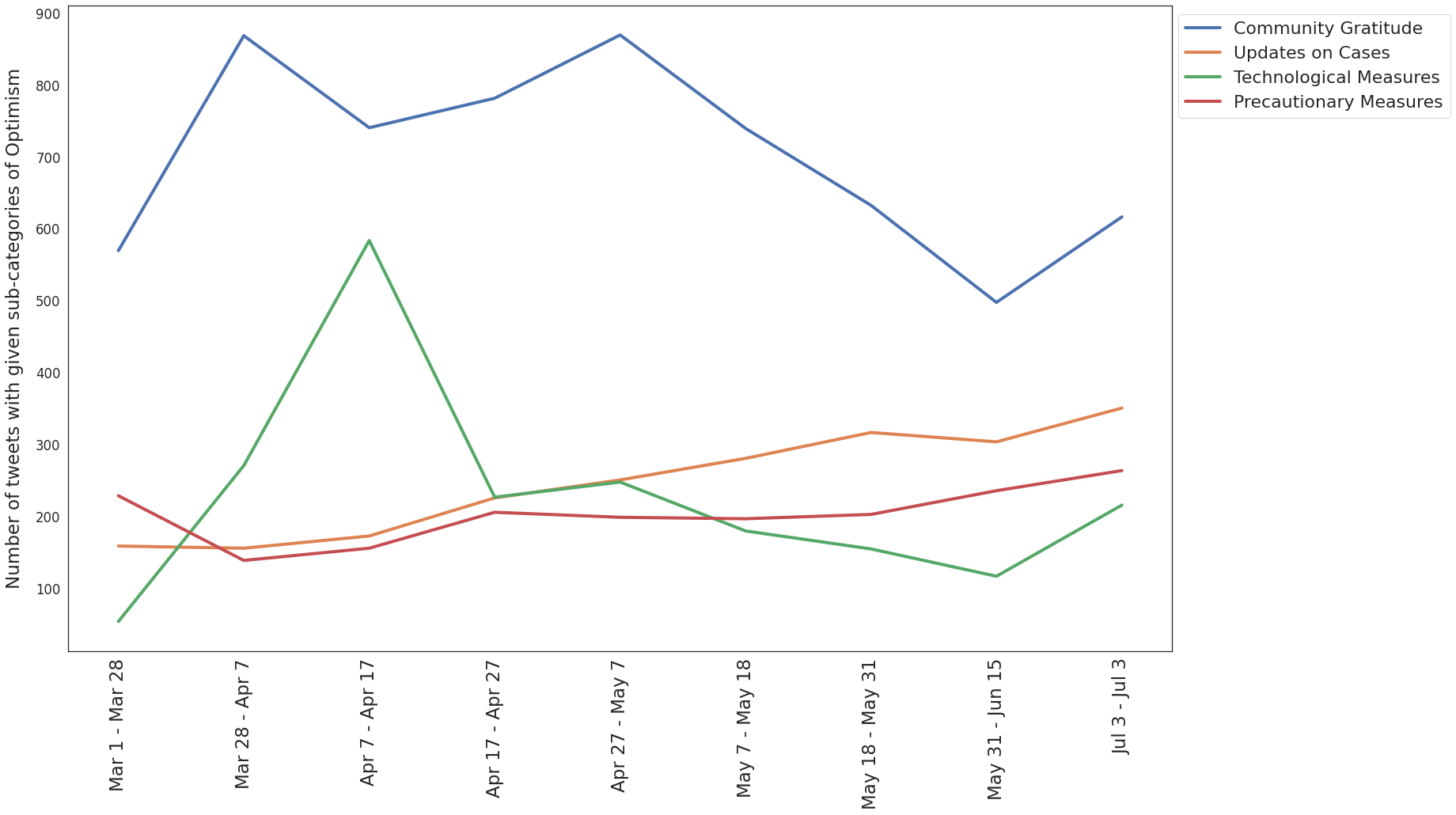}}
		\caption{Optimism}
		\label{fig:optimism_trend}
	\end{subfigure}
	\caption{Change in Subcategories of Emotional Triggers over time.}
	\label{fig:triggers}
\end{figure*}

In order to extract relevant factors or triggers affecting the public sentiments and emotions over time, we make use of an unsupervised autoencoder-based neural topic model \textit{ABAE} as proposed by Rudian He, et. al. in \cite{He2017AnUN}. We separately run the ABAE algorithm with the tweets of each of the six emotions captured in Table 4 and report their top contributing aspects. The extracted aspect terms for each emotion are filtered and assigned meaningful \textit{sub-categories} by means of a many-to-many mapping. In Figure \ref{fig:triggers}, we plot the trends of the subcategories for \textit{annoyed} (Fig. \ref{fig:annoyed_trend}) and \textit{optimism} (Fig. \ref{fig:optimism_trend} ). 

In Fig. \ref{fig:annoyed_trend}, we observe a clear peak from March 28th 2020 to April 8th 2020 due to the religious gatherings of a certain community, thereby triggering widespread criticism and hatred from the public. In Fig. \ref{fig:optimism_trend}, the plots show a high level of community gratitude in general, with occasional peaks which may be attributed to the events targeted at raising solidarity among the public. For the \textit{technological measures}, we see a gradual increase and a peak near the launch date of the Arogya Setu App - developed by the Indian Government to help our citizens in this pandemic.

\section{Conclusion and Future Work}\label{conclusion}
To summarize, we first build a multi-label emotion classifier that achieves state-of-the-art results on two benchmark datasets, \textit{AIT} (non-COVID) and \textit{SenWave} (COVID-specific). We then use our trained model to predict the emotions from India-specific tweets posted between March 1st 2020 and July 4th 2020. Using these predictions, we study the evolution of emotions among Indians towards the ongoing COVID-19 pandemic. We further study the development of salient factors contributing towards the changes in attitudes and draw interesting inferences. In future, we would like to extend our work with the \textit{Valence-Arousal-Dominance} or VAD model and take a multi-tasking approach to build our classifiers.

\bibliographystyle{ACM-Reference-Format}
\bibliography{main}
	
\end{document}